\documentclass[twocolumn]{emulateapj}
\usepackage{amsmath,amssymb,amstext}
\usepackage{color}
\usepackage{gensymb}
\usepackage[breaklinks,colorlinks,citecolor=blue]{hyperref}
\usepackage{float}

\usepackage{natbib}
\shorttitle{Atmospheric Models of Circumbinary Planets}
\shortauthors{May \& Rauscher}

\begin{document}

\title{Examining Tatooine: Atmospheric Models of Neptune-like Circumbinary Planets}
\author{E. M. May \& E. Rauscher}
\affil{University of Michigan}

\begin{abstract}
Circumbinary planets experience a time varying irradiation pattern as they orbit their two host stars. In this work, we present the first detailed study of the atmospheric effects of this irradiation pattern on known and hypothetical gaseous circumbinary planets. Using both a one-dimensional Energy Balance Model and a three-dimensional General Circulation Model, we look at the temperature differences between circumbinary planets and their equivalent single-star cases in order to determine the nature of the atmospheres of these planets. We find that for circumbinary planets on stable orbits around their host stars, temperature differences are on average no more than 1.0\% in the most extreme cases. Based on detailed modeling with the General Circulation Model, we find that these temperature differences are not large enough to excite circulation differences between the two cases. We conclude that gaseous circumbinary planets can be treated as their equivalent single-star case in future atmospheric modeling efforts.
\end{abstract}

\maketitle
\section{Introduction}
Circumbinary planets are a fascinating new regime of planets to be studied. While we have been drawn to them for decades in science fiction, it has only been in the past several years that such planets have begun to be discovered, leading to a new and exciting field of study. Orbiting two stars, circumbinary planets experience strong short-term variations in stellar irradiation over the course of their orbits owing to the motion of their host binary stars. Such variation is interesting when we begin to consider the possible effects on climate patterns for planets in these systems.
\par Binary stars are common in our galaxy. With the fraction of single stars with planets being at least 50\% \citep{Fress13}, we can say that planets are also a common occurrence in our galaxy. If we only consider planets which are 4 times the radius of Neptune (hereafter $R_{\mathrm{N}}$)\footnote{$R_{\mathrm{N}}$ = 3.88$R_{\mathrm{\oplus}}$ ; $R_{\mathrm{J}}$ = 11.2$R_{\mathrm{\oplus}}$}, the same study finds that just over 8\% of single stars are host to such objects. If circumbinary planets were to form at a similar rate, we could assume them to be just as common. In fact, \citet{ABUNDANCE_CB} finds that the fraction of binary stars with planets larger than 6 R$_{\oplus}$ is at least 10\% for coplanar circumbinary systems, a rate even higher than that for single-star systems. This makes circumbinary planets an extremely interesting field of study, since we can expect them to be quite numerous. 
\par As of writing, there have been 10 transiting circumbinary planets discovered around main sequence stars. The first was Kepler 16b \citep{KEP16b}. Following this were the discoveries of Kepler 34b and 35b \citep{KEP34_KEP35}, Kepler 47b and 47c \citep{KEP47}, Kepler 38b \citep{KEP38b}, Kepler 64b \citep{KEP64b_1,KEP64b_2}, Kepler 413b \citep{KEP413b}, Kepler 453b \citep{KEP453b}, and KOI 2939b \citep{KOI2939b}. With the exception of the most recent planet, which is Jupiter-sized, all are between approximately 0.75 - 2.25 $R_{\mathrm{N}}$. This agrees with studies which conclude that planets larger than 10 $R_{\mathrm{\oplus}}$ are uncommon around binary stars \citep{ABUNDANCE_CB}.
\par Many upcoming missions will provide opportunities to search for and discover new circumbinary planets. The Gaia mission is expected to discover between tens and hundreds of new circumbinary planets \citep{GAIA_CB}, depending on the abundance of giant circumbinary planets. In addition, microlensing has been shown as an effective way to discover circumbinary planetary candidates, with studies predicting that this is a sufficient method for candidate detection \citep{Microlens_CB}.
\par Atmospheric modeling is one of the most important ways we can gain insights into the circulation patterns of exoplanets. Through the use of models of varying complexities, we can learn about a planet's temperature structure and circulation and make predictions for observational signatures of various types of temperature structures due to, for example, shifted hot spots, seasonal effects, clouds, and composition. In this paper we study the atmospheres of circumbinary planets through the use of both a one-dimensional Energy Balance Model and a three-dimensional General Circulation Model in order to learn about the effects of flux variation on short timescales. Such a study for these types of planets has not been done previously, with the only similar study being of habitable zones around binary stars for hypothetical Earth-like planets \citep{Forgan13}.
\par By piecing together all of the information we have about the known set of circumbinary planets, and by studying them further through the use of models, we can begin to understand where such planets are similar to and different from their single-star counterparts. By modeling both a known circumbinary planet and a single-star planet with an equivalent constant irradiation, we then study the temperature and wind circulation patterns in the atmospheres of both planets in order to make comparisons between the two atmospheres. Through this work we are then able to answer the question of how a quickly varying stellar irradiation affects the atmosphere of a giant planet and make further statements as to the detectability of any differences between single-star and circumbinary planets.
\par In Section \ref{2_Method} we discuss the methods and models used in this work, with particular emphasis on model parameters. The results of our work, including various known and theoretical planets, is presented in Section \ref{3_Results}. Conclusions are given in Section \ref{4_Conclusions}.
\section{Method}
\label{2_Method}
We begin with a one-dimensional Energy Balance Model (Section \ref{EBM}) to study the general behavior of a planet's atmosphere over long periods of time. We use a three-dimensional General Circulation Model (Section \ref{GCM}) for more in depth studies of a specific planet in order to confirm results from the one-dimensional model. The benefits of using the Energy Balance Model (hereafter EBM) are that it is relatively simple, allowing us to model the atmosphere over many more planetary orbits and for a wide range of configurations. While the General Circulation Model (hereafter GCM) is beneficial in order to obtain more realistic and detailed three-dimensional results, it is much more computationally expensive. By using both models we are therefore able to obtain a better understanding of the effects of the varying irradiation pattern both in high detail and over a wider parameter space. We focus on Neptune-like planets, and all model parameters are derived based on this fact.
\subsection{Calculation of Orbits and Resulting Irradiation} 
\label{orbits}
\par We calculate orbits by assuming Keplerian motion, which is a good approximation over the short number of orbits we model. For the known circumbinary systems, observations over multiple planetary orbits support our choice for Keplerian motion. Both stars are placed on Keplerian orbits around their center of mass, with the planet on a Keplerian orbit around this same center of mass. Assuming the planet is approximately Neptune-like in mass, it is not massive enough to significantly perturb the stellar orbits away from Keplerian motion over the timescales we consider. Our assumption of planet mass is based on the measured radii of known circumbinary planets, and mass-radius models and observations which do not predict a planet of this size to be dense enough to be terrestrial  \citep{{MR_3},{MR_2},{MR_1}}. 
\par
\begin{figure}[ht]
\epsscale{1.35}
\plotone{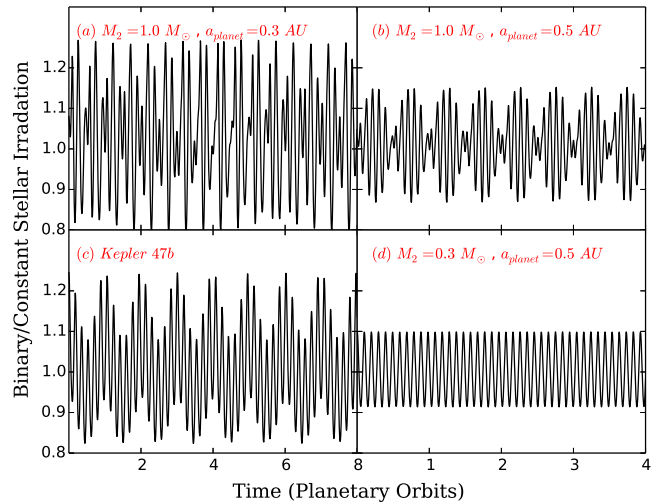}
\caption{Flux variation for several different systems. For (a), (b), and (d), The primary stellar mass is set to 1 M$_{\odot}$ and the stellar separation is set to 0.1 AU (see section \ref{grid}) with zero eccentricity for all orbits. Plot (c) shows the irradiation variation for the planet Kepler 47b (see Table \ref{47b_data} for system parameters). Note that for planets on close orbits around equal mass binary systems (plot a) the irradiation pattern has both a higher amplitude and is more irregular than for planets on more distant orbits around a system mostly dominated by one star (plot d). For each system, the overall variation in flux is due to the motion of the binary stars relative to the location of the planet. For the case of Kepler 47b (plot c), there is a secondary super-imposed long period variation due to the planet's slight eccentricity. }
\label{FluxCurves}
\end{figure}

 For all known systems, the irradiation pattern is non-repeating over timescales of both the stellar orbits and the planetary orbits. General patterns do emerge which depend on the location of the more luminous star or stars relative to the planet. As shown in Figure \ref{FluxCurves}, planets with more massive secondary stars (top two panels) experience a greater variation in their stellar irradiation pattern over a given planetary orbit, while those with lower mass secondaries (bottom two panels) have stellar irradiation patterns which become much simpler, beginning to resemble simple sinusoidal curves. Further, planets on longer period orbits experience less extreme irradiation patterns (smaller amplitude variations) as the motions of the stars become less important simply due to the larger distance between the planets and the stars. This can also be seen in Figure \ref{FluxCurves} when comparing the left two panels to the right two panels.
\subsection{Energy Balance Model} 
\label{EBM}
Historically, EBMs have been used to predict the long-term climate of terrestrial-like planets  \citep[see][]{North81,Forgan13,Vladilo15}. EBMs are designed to be relatively simple, yet complex enough to encapsulate the relevant physics of atmospheric heating. The simplicity of these models comes from their single dimension in space, latitude. For terrestrial planets, models predict the planet's surface temperature - so there is no need to study the various heights in the atmosphere. EBMs calculate the temperature at each point in latitude space, evolving it forward in time to represent the evolution of the surface or atmospheric temperature over a period of time
\par Previous works have been applied to Earth-like planets only - here we make modifications to account for the thicker atmosphere of Jovian-like planets. When modeling planets dominated by a thick atmosphere, we  decide to modify the coefficients in such a way that we are integrating over the range of pressures in the atmosphere in which heating takes place. This lets us gain insight into the temperature of the thermal (infrared emission) photosphere of the planet. This choice becomes important as we discuss observables in section \ref{obs}. 
\par The general form of the EBM is given by
\begin{equation}
\label{EBMeq}
C\frac{\partial T}{\partial t} -\frac{\partial}{\partial x}\left(D\left(1-x^2\right)\frac{\partial T}{\partial x}\right)=S\left(1-A\right)-I.
\end{equation}
Here $x\equiv\sin\varphi$ is the single dimension in the model, with $\varphi$ being latitude. $T$ is the local temperature at some time. C is the atmosphere's heat capacity per unit area; D is the diffusivity of the atmosphere, S is the stellar irradiation, A is the atmosphere's albedo, and I is the cooling function of the atmosphere. In the next few sections we discuss all parameters and their physical roles in the model, as well as our modifications to account for the specific cases we study. The adopted values are listed in Table \ref{EBMparams}. 
\subsubsection{C: Heat Capacity Per Unit Area} 
As discussed in the introduction, most of the currently known circumbinary planets are approximately Neptune-sized, ranging from 0.768 $R_{\mathrm{N}}$ (Kepler 47b, \cite{KEP47}) to 2.20 $R_{\mathrm{N}}$ (Kepler 34b, \cite{KEP34_KEP35}), with a single outlier at 1.52 $R_{\mathrm{J}}$ (KOI 2939b, \cite{KOI2939b}) . For this reason, we chose to use a Neptune-like composition for all planets modeled in this work. This approximation allows better comparison across systems by minimizing free parameters.
\par Previous work on circumbinary planets using EBMs \citep{Forgan13} focuses on hypothetical Earth-like planets. Here we seek to model the known planets, with much of their mass and radius being dominated by atmosphere. Because of this, we do not need to take into account the fraction of the planet covered in ice, water, and solid surfaces as is often done with EBMs. Therefore, our heat capacity is calculated based on our choice of atmospheric composition with a mean molecular weight of 2.5 g/mol \citep{PSC}, a value commonly used for Neptune. This corresponds to a specific gas constant of 3.2$\times$10$^7$ erg g$^{-1}$ K$^{-1}$, which we can covert to heat capacity per unit area at the pressure level (thermal photosphere, 0.12 bar) we are studying. $C$ is then set to 4.154$\times$10$^{10}$ erg cm$^{-2}$ K$^{-1}$.
\subsubsection{D: Diffusivity} 
\label{Diff_EBM}
The diffusivity coefficient ($D$) encapsulates the atmospheric transport of heat from the equator towards the cooler poles. For Earth-like planets, the diffusivity coefficient is set based on empirically fitting models to reproduce the observed temperature structure on Earth. In a similar way, we can use more sophisticated models of gas planets to predict how meridional heat transport occurs. This is discussed in depth in section \ref{GCMdiff}. The diffusion constant could in principle be set as a function of $x$ (latitude), but we use a constant value in this work, which we find is able to capture the relevant physics without adding more unknowns into the model. 
\par Using results of a preliminary GCM (model details are discussed in Section \ref{GCM}), we obtain an initial estimate of the diffusion coefficient based on the rate of energy transport in this preliminary model. This served as a base value which was then varied over multiple runs of the EBM in order to obtain a latitudinal temperature structure which corresponded to the results of the GCM. As a result, we use a diffusion coefficient of 4$\times$10$^3$ erg cm$^{-2}$ s$^{-1}$ K$^{-1}$. For comparison, typical values of diffusivity used for Earth-like planets are of order $10^2$ erg cm$^{-2}$ s$^{-1}$ K$^{-1}$.
\par It is of importance to note that this value is the least constrained of the parameters used here. Small variances have little affect on the resulting temperature structures. However, extremely large values of the diffusion coefficient (greater than of order 10$^4$ erg cm$^{-2}$ s$^{-1}$ K$^{-1}$) correspond to efficient heat transport to the poles, where it is then radiated out to space, leading to an global cooling effect. Similarly, extremely low values of the diffusion coefficient (smaller than of order 10$^2$ erg cm$^{-2}$ s$^{-1}$ K$^{-1}$) correspond to heat being trapped near the equator, leading to an global heating effect.
\subsubsection{S(1-A): Stellar Irradiation and Albedo} 
\label{SandA_EBM}
Stellar irradiation ($S$) is set by the motions of the binary stars in the system. The time ($t$) dependence of this parameter is determined by the specific planetary and stellar orbits for any given system. The spatial ($x$) dependence of this parameter is rooted in the fact that the equator receives more irradiation than the poles for zero (or very low) obliquity systems, represented by a factor of cosine of latitude in the received irradiation. Based on the small angular separation of the two stars in the planet's sky, we do not need to account for two separate substellar points, and instead treat them as one point, again letting us make the claim that the equator is heated more than the poles.
\par As discussed in Section \ref{orbits} and demonstrated in Figure \ref{FluxCurves}, the binary mass fraction and the planet's semi-major axis play a large role in the stellar irradiation. Additionally, we ignore the effects of stellar eclipses, as they occur on time scales which are short relative to the radiative timescale of the atmosphere.
\par Albedo ($A$), or the amount of light reflected off the top of the atmosphere, is set to 0.3. This value has been used for Neptune previously (such as in \cite{LiuSch10}), so we find it appropriate to use due to other parameters being set to Neptune values as well.
\subsubsection{I: Cooling Function} 
The cooling function determines the planet's outgoing radiation for any given point in latitude. Again, we examine results from the preliminary GCM to determine the appropriate amount of outgoing radiation necessary to produce the temperature structure seen in this more complicated model. As the cooling function represents the outgoing flux, we can represent it as
\begin{equation}
I(T)=\frac{\sigma T^4}{\alpha},
\end{equation}
where $\alpha$ is a constant related to the optical thickness of the atmosphere, determined by the GCM. Our initial parameter guess from the preliminary GCM reproduces the expected temperatures in the EBM without any further fine tuning. Based on results for outgoing radiation and temperature structure from the GCM, an $\alpha$ value of 1.682 is used.
\begin{deluxetable}{cc}
\tabletypesize{\small}
\tablecaption{Energy Balance Model parameters \label{EBMparams}}
\tablewidth{0.5\textwidth}
\tablehead{
\colhead{Parameter} & \colhead{Value Adopted}}
\startdata
Heat Capacity, C & 4.154$\times$10$^{10}$ erg cm$^{-2}$ K$^{-1}$ \\ 
Diffusivity Coefficient, D & 4$\times$10$^3$ erg cm$^{-2}$ s$^{-1}$ K$^{-1}$ \\ 
Albedo, A & 0.3 \\ 
Cooling Function, I & $I(T)=\sigma T^4 / \alpha$ \\
Cooling $\alpha$ & 1.682 
\enddata
\end{deluxetable}
\subsection{General Circulation Model} 
\label{GCM}
\par GCMs are three dimensional climate models which calculate temperature and winds at every point in latitude, longitude, and pressure space using basic information about the planet we seek to model - such as atmospheric composition, solar irradiation, size, and rotation rate. Because of their complexity, these models can be used to predict observational signatures in the form of infrared radiation for a wide array of planet types and provide a base for detailed study of planetary atmospheres.
\par We use the GCM detailed in \cite{RM12}, with a modification to account for the variation in stellar irradiation, calculated as discussed in section \ref{SandA_EBM} and \ref{orbits}. The model is built upon the primitive equations of meteorology, which are a standard reduction of the Navier-Stokes equations including assumptions of inviscid flow, vertical hydrostatic equilibrium, and small vertical flow and scales relative to the horizontal components. See \cite{Vallis2006} for further explanation. Heating is treated using double gray radiative transfer such that the incoming radiation (optical) and the outgoing radiation (infrared) each have their own absorption coefficient. For a more detailed description of the model, see \cite{RM12} and sources within. See Table \ref{GCMparams} for a list of parameters used for a Neptune-like planet.
\begin{deluxetable}{cc}
\tabletypesize{\small}
\tablecaption{General Circulation Model Parameters \label{47bprop}}
\tablewidth{0.5\textwidth}
\tablehead{
\colhead{Parameter} & \colhead{Value Adopted}}
\startdata 
Planet density, $\rho_{\mathrm{N}}$ & 1.64 g/cm$^3$ \\
Rotation rate, $\Omega_{\mathrm{N}}$ & 1.08$\times$10$^{-4}$ s$^{-1}$ \\
Specific gas constant, R & 3.2$\times$10$^7$ erg g$^{-1}$ K$^{-1}$ \\
Optical absorption, $\kappa_{vis}$ & 8.14$\times10^{-4}$cm$^2$g$^{-1}$ \tablenotemark{a}\tablenotetext{a}{Adapted to our model formulation from \cite{LiuSch10}}\\
Infrared absorption, $\kappa_{IR}$ & 1.49$\times10^{-2}$cm$^2$g$^{-1}$ \tablenotemark{a} \\ 
Internal heat flux, F$_{int}$ & 0.433 W m$^{-2}$ \tablenotemark{a} \\
Irradiated flux, F$_{irr}$ & Varies - Section \ref{SandA_EBM} 
\enddata
\tablecomments{Values with a subscript N are set to accepted parameters for Neptune. We chose density as the constant parameter across various Neptune-like planets. The radius of the each planet is set from observations, and density is used to obtain a mass and gravitational acceleration for use in the General Circulation Model.}
\label{GCMparams}
\end{deluxetable}
\subsubsection{GCM Heat Transport - applications to the EBM} 
\label{GCMdiff}
Using equations adapted from \cite{ClimateBook}, we can relate the heat transport in the EBM to the heat transport in the GCM by writing
\begin{equation}
\Phi\equiv - D\frac{\partial T}{\partial \varphi},
\end{equation}
where D is the diffusion coefficient defined in section \ref{Diff_EBM}. $\Phi$, given by 
\begin{equation}
\Phi=\frac{1}{R_p}\int_0^{P_s}v\left(c_pT+gz\right)\frac{dP}{g}
\end{equation}
relates to the rate of energy transport across latitude bands, calculated as 2$\pi R_p^2\Phi\cos\varphi$. In all of the above, $R_{\mathrm{p}}$ denotes the radius of the planet, $P$ the pressure of a given level, $P_{\mathrm{s}}$ the representative bottom boundary pressure, $T$ the temperature at that pressure level, and $g$ the gravitational acceleration. Using output of the GCM, $\Phi$ can be computed for a planet of our chosen composition as a function of latitude, $\varphi$. We can then compute the diffusion coefficient $D$ as a function of latitude. For simplicity, we adopt an average value, which as discussed above, has been determined to be sufficient for 1D models 
\section{Results}
\label{3_Results}
In the following sections we study the planet Kepler 47b using both the EBM and the GCM. First, by running the one-dimensional EBM, we obtain limits on the atmosphere's temperature variations due to the time varying flux. By then running the three-dimensional GCM, we can further study how these variances may or may not affect the planet's circulation. 
\par We then extend our results to all known circumbinary systems as well as a grid of hypothetical circumbinary systems in order to thoroughly examine which regions of parameter space are host to planets which are most affected by their orbit around a binary star system. To compare our results to a single-star system, we calculate an equivalent single-star case in which the total luminosity of both stars is held unmoving at the center of mass. This allows for a direct comparison between a single-star case and a circumbinary case with all other variables the same. 
\subsection{Kepler 47b} 
Kepler 47 is a multi-planet binary system, with two confirmed Neptune-sized planets \citep{KEP47}. We initially decided to study Kepler 47b in the most detail due to a comparable radiative time scale of the atmosphere and timescale over which the irradiation is varying (similar to the orbital period of the stars). The radiative timescale describes how quickly or slowly a planet's atmosphere is able to respond to changing environmental conditions, therefore a planet which has a radiative timescale nearly that of the short scale changes in irradiation may have interesting coupling between the variation in stellar heating and the atmospheric response. Atmospheres with longer radiative timescales will be unaware of the varying flux, and we can expect them to then respond to the average irradiation, while those with shorter radiative timescales will respond very quickly to changes in stellar flux. System parameters are given in Table \ref{47bprop}. 
\par The irradiation pattern for Kepler 47b is shown in the bottom left panel of Figure \ref{FluxCurves} and is fairly regular with two superimposed sinusoidal patterns. The shorter of the two periods is due to the motions of the host stars, and the other, longer period, sinusoidal pattern is due to the slight eccentricity of the planetary orbit. This eccentricity is set to the upper limit given in \citet{KEP47} in order to test the combined effect of eccentricity and binary motions. 
\begin{deluxetable}{cccc}
\tabletypesize{\small}
\tablecaption{Kepler 47 System Parameters \label{47bprop}}
\tablewidth{0.5\textwidth}
\startdata 
\cutinhead{\textbf{Stellar Properties}} 
Parameter & Primary Star & Secondary Star \\ [3pt] \hline
Mass (M$_{\odot}$) & 1.043 & 0.362  \\
Radius (R$_{\odot}$) & 0.964 & 0.3506 \\
Temperature (K) & 5636 & 3357 \\
\cutinhead{\textbf{Stellar Orbit}}
Stellar Separation (AU) & 0.0836 \\
Orbital Period (Earth Days) & 7.448 \\
Eccentricity& 0.0234 \\
\cutinhead{\textbf{Planet b Properties}}
Radius (R$_{Nept}$) & 0.767 \\
Semimajor Axis (AU) & 0.2956 \\
Orbital Period (Earth days) & 49.514 \\
Eccentricity & \textless 0.035
\enddata
\tablecomments{All values have been adopted from \cite{KEP47}. We chose to adopt the upper limit for the planet's eccentricity in order to maximize any variations.}
\label{47b_data}
\end{deluxetable}
\subsubsection{Kepler 47b - EBM} 
Using the EBM, we run both the circumbinary case with flux varying according the motions of the stars relative to the planet, and a single-star case set up as described above. Both models run for 20 planetary orbits (where one orbit is 49.5 Earth days, see Table \ref{47bprop}), which serves to give us a long enough base-line for comparison between the two cases. The atmosphere equilibrates over the first planetary orbit, so for this reason we do not include the first orbit in our analysis. 
\begin{figure*}[ht]
\epsscale{1.4}
\centering
\plotone{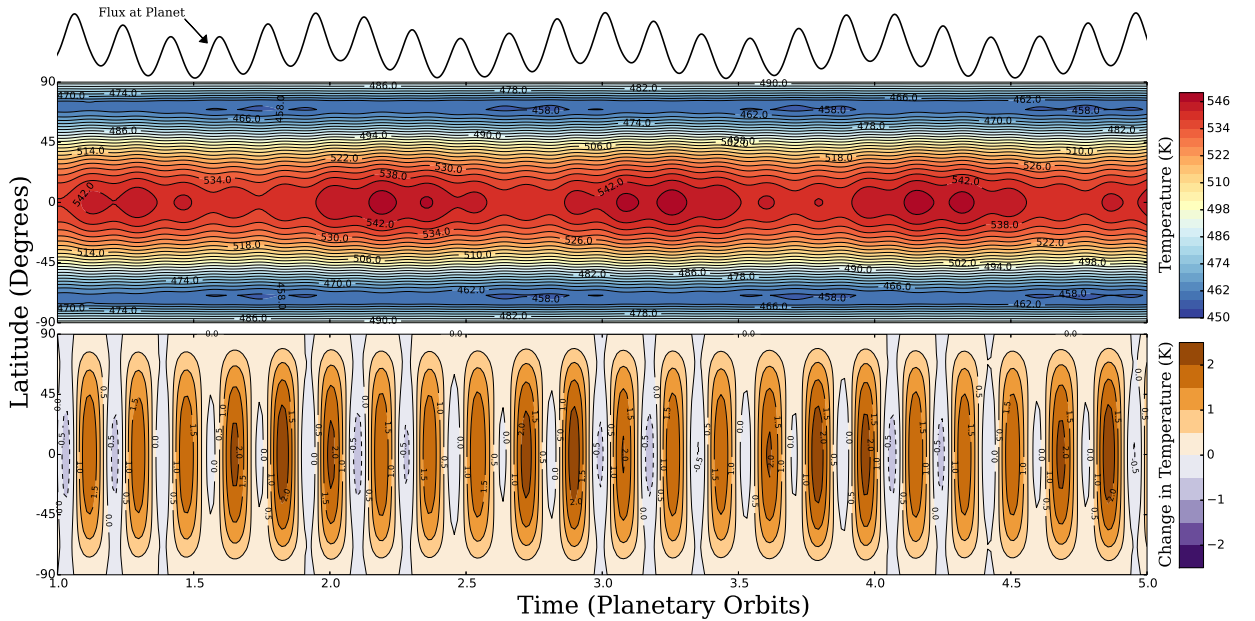}
\caption{Results of the Energy Balance Model for the circumbinary planet Kepler 47b. The top curve represents the stellar irradiation pattern received by the planet over the course of the 4 planetary orbits plotted. The bottom two plots represent the evolution of the thermal photosphere's temperature over time. For the middle plot, the absolute temperature is plotted. We see peaks in temperature at appropriate delays after peaks in irradiation due to the time lag of the atmosphere. The bottom plot shows the relative changes in temperature as compared to the equivalent single-star case for Kepler 47b. We see maximum positive and negative changes of 2 Kelvin, implying that the overall effect of the binary stars is minimal for this case.}
\label{EBM_47b}
\end{figure*}
As the EBM is a one-dimensional model, the output represents the temperature across all latitude bands evolved forward in time. Shown in the middle frame of Figure \ref{EBM_47b} is the model output at the thermal (outgoing) photosphere in the circumbinary case through the fifth planetary orbit. We find that peaks in temperature correspond to peaks in irradiation after accounting for the expected delay due to the radiative lag of the atmosphere. For this system, the radiative timescale corresponds to approximately 0.14 planetary orbits, or approximately 7 days.
\par As shown, times of peak temperature are only a few Kelvin warmer than other times. This is reasonable, as we can not expect a direct comparison between changes in irradiation and atmospheric temperature because the atmosphere dampens out the irradiation changes. A direct comparison between flux and temperature ($\Delta F\rightarrow \sigma\Delta T^4$) suggests changes of order tens of Kelvin, so our results of a few Kelvin is reasonable.
\par By eye, we can see that the circumbinary planet shows variations we would not expect for planets receiving a constant amount of irradiation over the course of its orbit. We would like to quantify how different a circumbinary planet is from its equivalent single-star case, so we define a parameter $\eta$ representing the mean of the absolute value of the fractional difference in temperature for the circumbinary case as compared to the single-star case.
\par For Kepler 47b, we find an $\eta$ of 0.2\%. The bottom frame of Figure \ref{EBM_47b} shows the temperature deviations over the first through fifth planetary orbits as compared to the single-star case, demonstrating that for the case of Kepler 47b, the maximum deviation away from the single-star model is 6K, which is less than 1\%. 
\subsubsection{Kepler 47b - GCM} 
\begin{figure}
\epsscale{1.0}
\plottwo{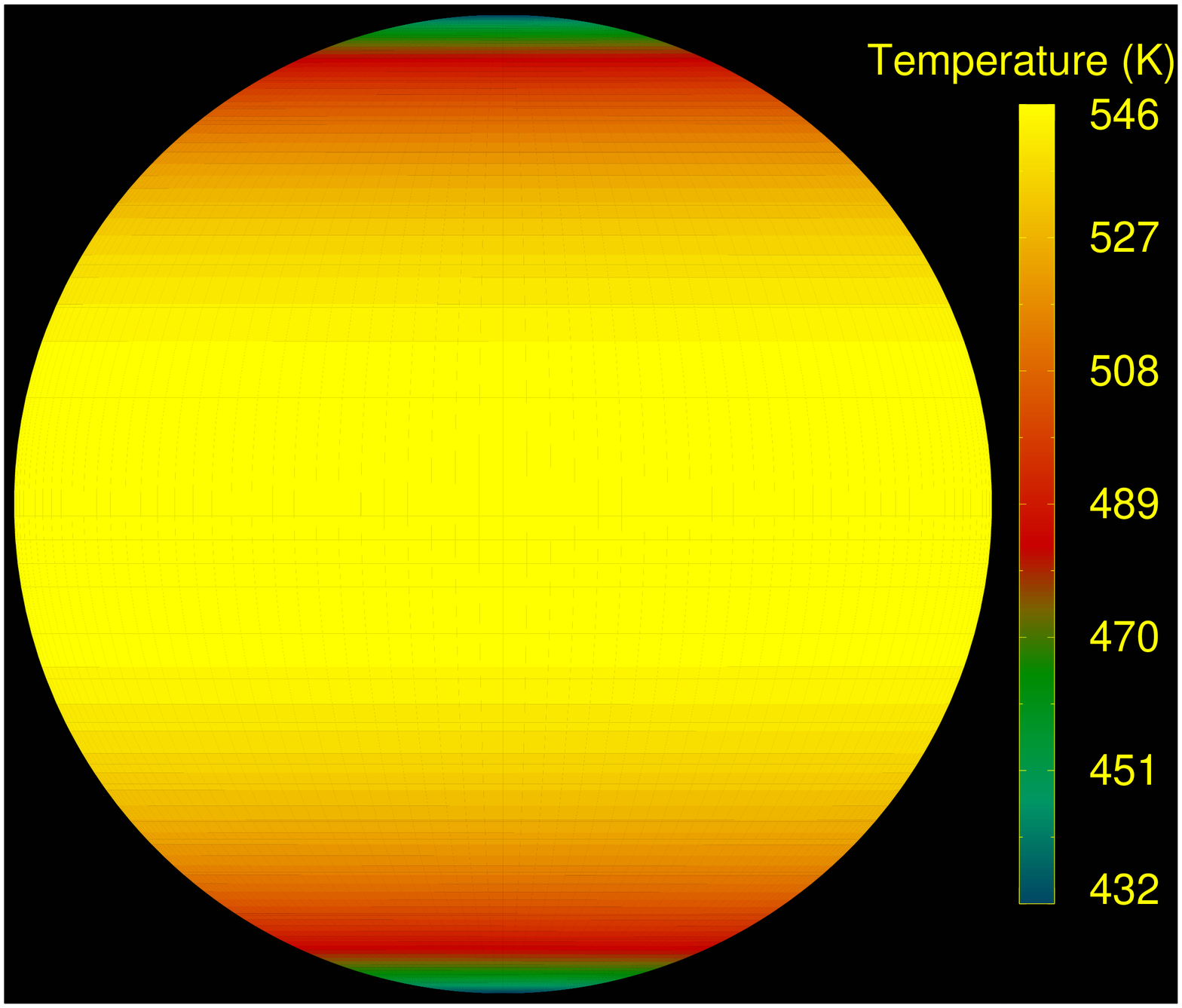}{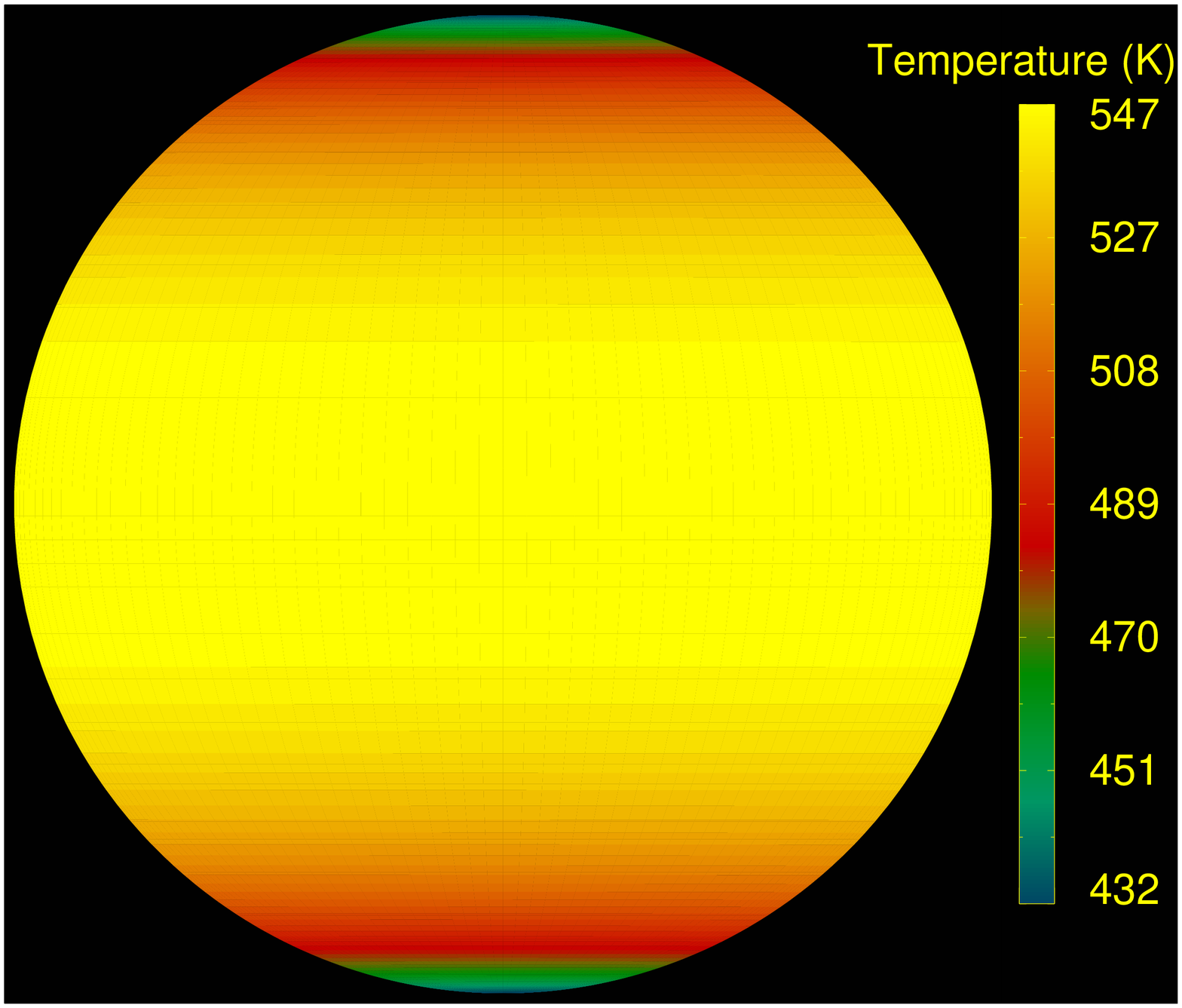}
\plottwo{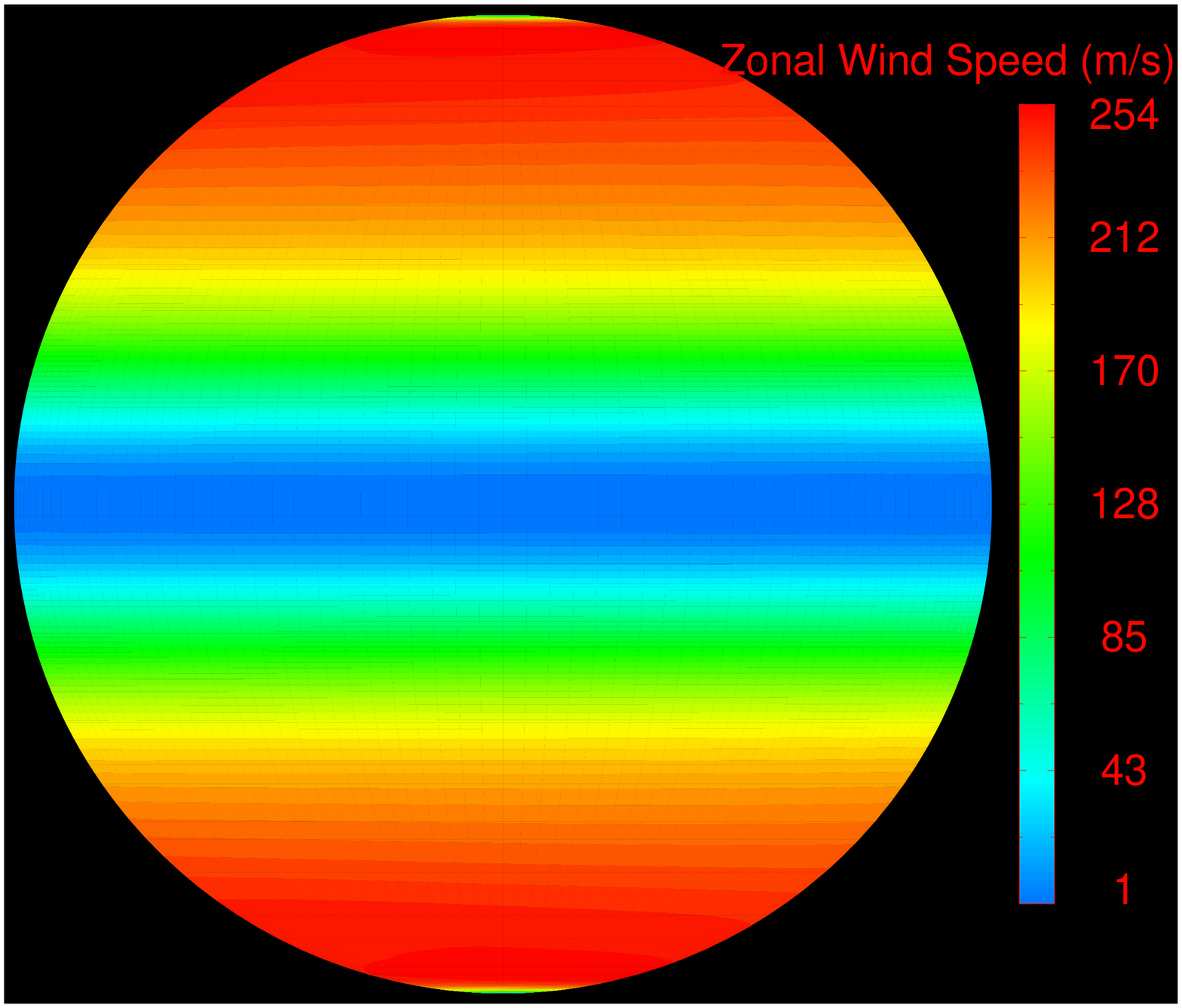}{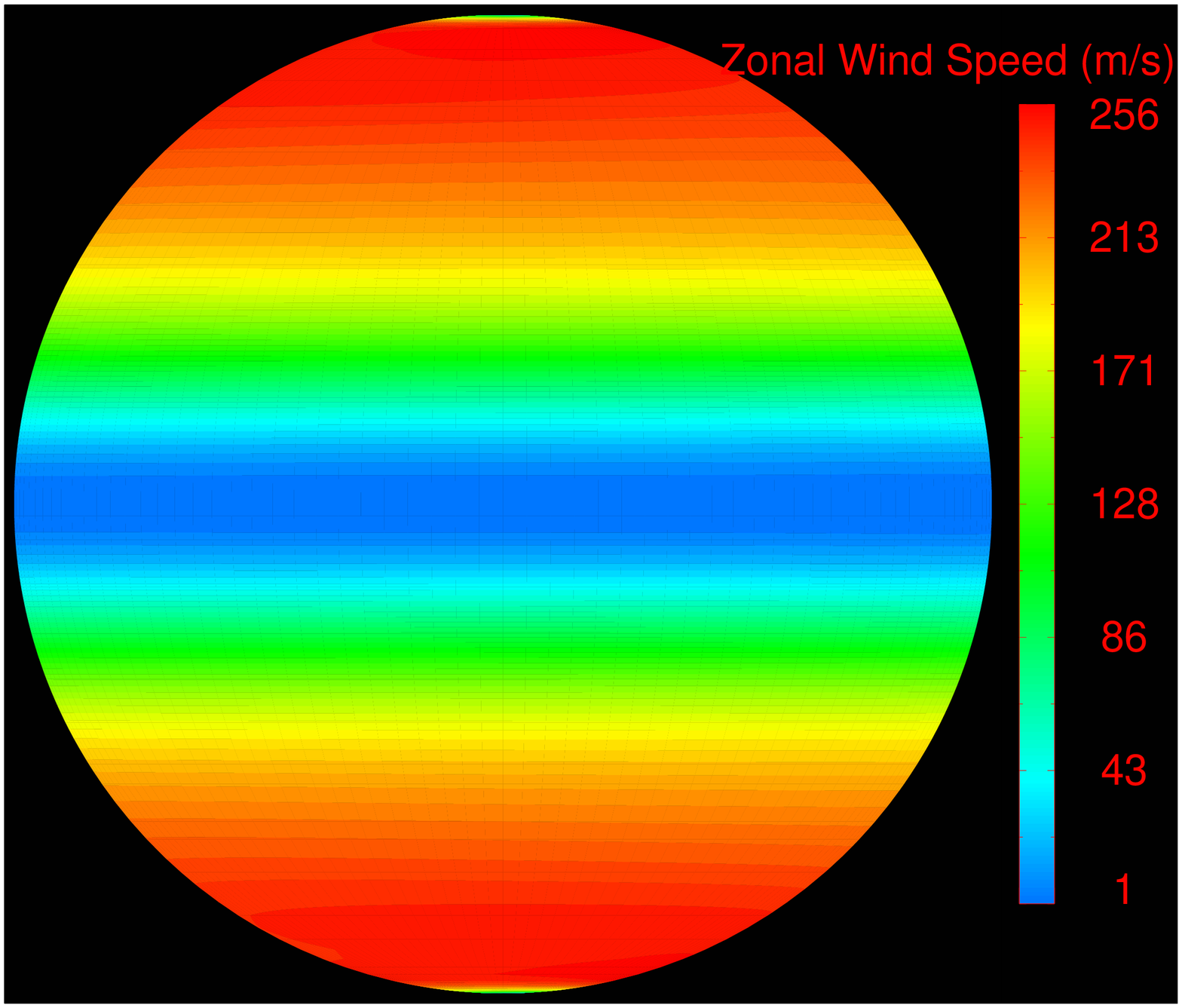}
\caption{GCM results for the circumbinary case (left) and the single-star case (right) at a single point in time for Kepler 47b. The top panel shows the temperature of the thermal photosphere, and the bottom panel shows the east-west winds at the thermal photosphere. We see no discernible difference between the circumbinary case and the single-star case.}
\label{GCMplots}
\end{figure}

In order to determine the magnitude of these temperature variations on the planet's circulation, we study this planet with the full three-dimensional GCM. We do not expect that a mean deviation of 0.2\% is large enough to excite noticeable changes in the planet's circulation, however, we explore the possibility here for completeness. Figure \ref{GCMplots} shows the full three-dimensional model results for both the circumbinary case (left) and the single-star case (right). The top row shows the temperature of both cases at a snapshot in time, and the bottom row shows the zonal (East-West direction) winds for both cases at the same point in time.
\par GCM results suggest differences in temperatures and wind speeds for the circumbinary case compared to the single-star case which are less than 1 K and 1 m/s, respectively, at any given point in time. We find that for $\eta\approx$ 0.002 (as calculated from the EBM), the irradiation pattern due to the binary motion does not cause noticeable changes in planetary circulation as compared to the equivalent single-star case, unsurprisingly.
\par
\begin{figure}
\epsscale{1.2}
\plotone{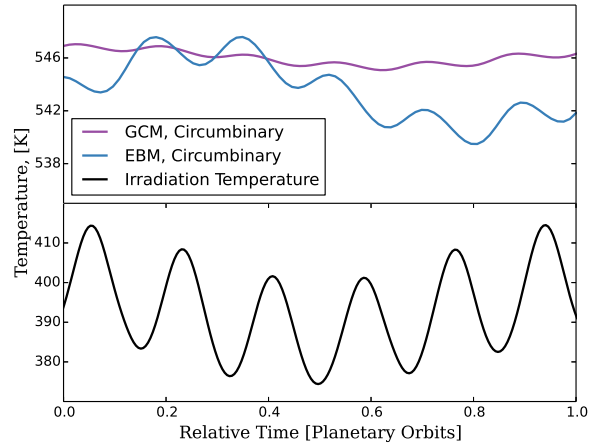}
\caption{Temperature predictions from the General Circulation Model (Purple), Energy Balance Model (Blue), and Irradiation Temperature (Black) for Kepler 47b. We see smaller amplitudes in temperature variation as we move from the least complicated prediction (black body) to the most complex model (general circulation) suggesting that the atmosphere is efficient at dampening out the irradiation pattern. Note that peaks between the black body and other models are offset due to the delayed response of the atmosphere.}
\label{EBM_GCM}
\end{figure}

 A further comparison of the zonal-averaged temperature at the equator from the GCM (an average temperature across the latitude band sitting at the equator), the equator temperature of the EBM, and the irradiation temperature at the equator (T=$(S(t)/\sigma)^{(1/4)}$) over the course of one orbit shows that the GCM produces temperature variations of even smaller amplitudes than those found in the EBM (Figure \ref{EBM_GCM}). Comparing both to the amplitudes of the temperature variations we could expect if the atmosphere responded immediately to changes in irradiation, we see that the atmosphere in both models does indeed dampen out the irradiation changes. The results of this comparison allows us to determine that any atmospheric effects due to the irradiation pattern exhibited in the Kepler 47 system are negligible and become even less pronounced when we use the GCM, which more correctly models the heating and cooling of the atmosphere as compared to the EBM.
\subsection{Expanding Models to More Planets} 
\label{grid}
\begin{figure}
\epsscale{1.0}
\plotone{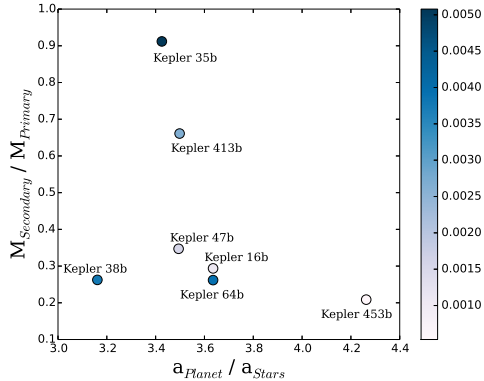}
\caption{$\eta$ values for 7 of the 10 known circumbinary systems as of writing. Kepler 34b, Kepler 47c and KOI2939b are excluded due to their large semi-major axes, placing all three off the right of the plot. All have $\eta$ values less than 0.1\%.}
\label{FracDiff_known}
\end{figure}

We seek to draw a conclusion for a wider range of circumbinary planets, and so we start by looking at the 10 circumbinary planets which have been confirmed. For relevant orbital parameters, see sources  \cite{KEP16b} (Kepler 16b), \cite{KEP34_KEP35} (Kepler 34b and 35b), \cite{KEP47} (Kepler 47b and 47c), \cite{KEP38b} (Kepler 38b), \cite{KEP64b_1,KEP64b_2} (Kepler 64b), \cite{KEP413b} (Kepler 413b), \cite{KEP453b} (Kepler 453b), and \cite{KOI2939b} (KOI2939b). For each system, we run circumbinary-case models and single-star models for 20 orbits and calculate their $\eta$ value using the EBM. Resulting $\eta$ values are shown in Figure \ref{FracDiff_known}.
\par Due to each system's unique orbital parameters and host stars, planets such as Kepler 16b and Kepler 64b can lay close to each other in the shown parameter space, but experience different effects due to their binary stars. In this case, both of the stars in the Kepler 64 system are at least twice as massive as those in the Kepler 16 system, causing the stars to move through their orbits much faster. This causes the variations in irradiation to be much more extreme (see Figure \ref{FluxCurves} for a demonstration of the variation in flux across different systems). 
\par Here we also see that Kepler 35b experiences the most extreme variations over the single-star case, at an $\eta$=0.0051, or 0.51\%. This matches our expectations that this planet should experience the most variations due to its closer proximity to two nearly equal mass stars.  Figure \ref{FluxCurves} demonstrates this through the differences in irradiation patterns for stars near and far from equal mass stars. Although not shown in Figure \ref{FracDiff_known}, Kepler 47c and KOI 2939b experience the smallest variations with $\eta$=0.00013 (0.013\%) and 0.00088 (0.088\%) respectively due to their large orbital distances of 0.989 AU and 2.72 AU.
\par Though there are only 10 announced circumbinary planets as of writing, we also wish to expand these results to a wider array of possible circumbinary planets. To do so, we develop a grid of  hypothetical planets with a set stellar separation of 0.1 AU (corresponds to slightly less than the average stellar separation for the known systems), a set primary star mass of 1.0 $M_{\odot}$ (corresponds to the average primary star mass of the known systems), and then vary the planetary semi-major axis and the secondary star mass. Further, because we are only focused on studying circumbinary planets orbiting two main-sequence stars, we can apply main sequence scaling relations based on a fully radiative approximation to convert stellar mass to luminosity,
\begin{equation}
\frac{L}{L_{\odot}}=\left(\frac{M}{M_{\odot}}\right)^{10.1/2}
\end{equation}
and update the orbital periods of the stars and planet as necessary by Kepler's laws with P=$2\pi\sqrt{a^3/(G(M_1+M_2))}$. While low mass stars are generally fully convective, and this is likely to then over predict the luminosity of such a star, low mass stars will contribute minimally to the total luminosity of the system, and as such it is sufficient to simplify the general set of mass luminosity relations to one relation for all stars. Additionally, all orbits are set to zero eccentricity so that the only variation produced is due to the motions of the binary stars, and not dependent on orbital eccentricity. 
\par
\begin{figure*}[ht]
\epsscale{1.05}
\plotone{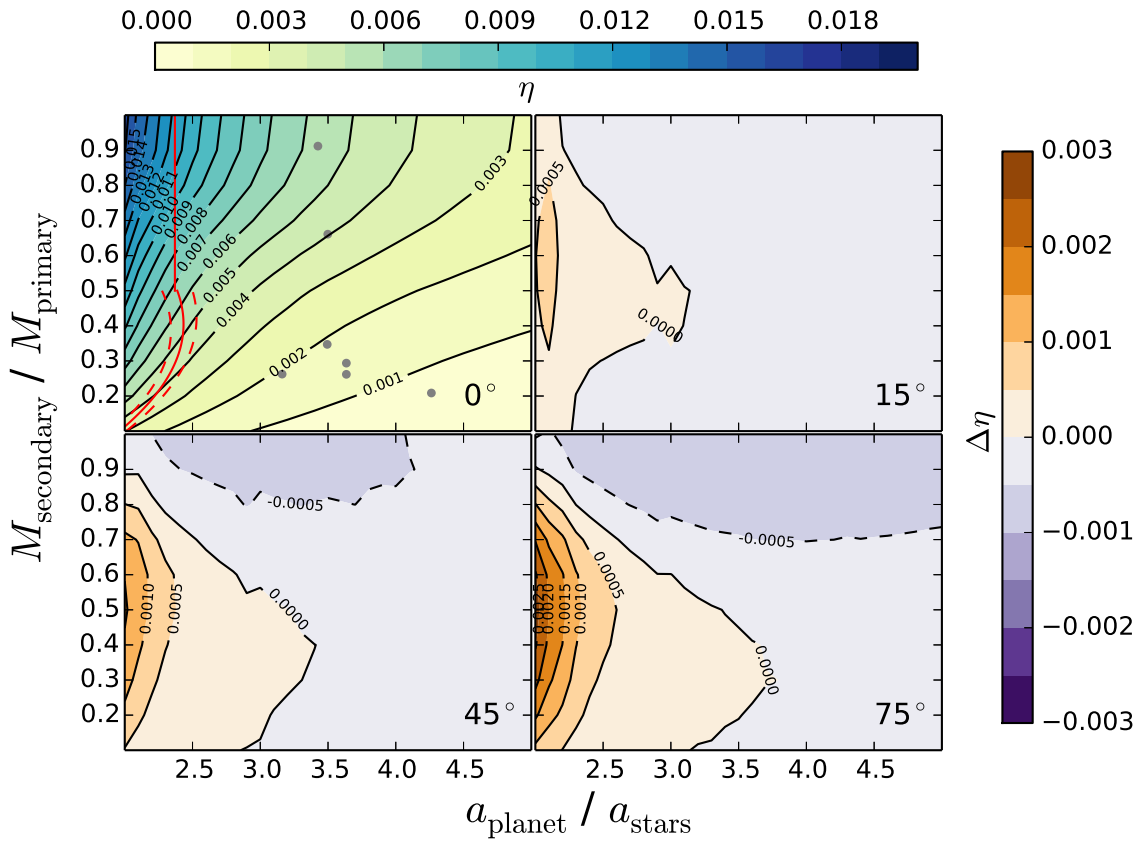}
\caption{$\eta$ values for the grid of models with $M_{\mathrm{primary}}$ and $a_{\mathrm{stars}}$ held constant at 1.0 $M_{\mathrm{\odot}}$ and 0.1 AU respectively. The top left represents no seasons, or 0 degrees obliquity. The red curve gives the approximate limit of stability as established in \cite{stable, stable89} with appropriate errors. Grey dots represent the planets shown in Figure \ref{FracDiff_known}. Each other frame represents a different obliquity, corresponding to stronger seasons, where the results of the zero obliquity case have been subtracted away to demonstrate which regions of parameter space are most affected by the introduction of seasons. As seasonal variation is made stronger due to an increased axial tilt, we see that planets far away from a system with an equal mass ratio become more like their single-star comparisons (shifts towards smaller $\eta$ values) and planets close in to nearly all types of mass ratios experience more variation over their single-star comparisons (shifts towards larger $\eta$ values), though it is noted that these fall within the region of instability and are therefore physically unimportant.}
\label{Grid}
\end{figure*}

 We then run a grid of planets for 0.2 AU \textless $a_{\mathrm{p}}$\textless 0.5 AU and 0.1 $M_{\mathrm{\odot}}$ \textless $M_{\mathrm{2}}$\textless 1.0 $M_{\mathrm{\odot}}$ to examine the ranges where atmospheric variation over the single-star case may become important. For each system, we calculate its $\eta$ value. Results are shown in the top left of Figure \ref{Grid}. We see the expected trend of planets on close orbits around a binary system of equal-mass stars exhibiting a greater variance over its equivalent single-star case (larger $\eta$), with planets far away from systems dominated by one of the stars being much more similar to their equivalent single-star case (smaller $\eta$). 
\par We note that the systems close into two equal-mass stars are unlikely to be stable (see \cite{stable}) but we find it informative to model these systems regardless. From \cite{stable}, we take the limit of stability (a$_{\mathrm{critical}}$) for zero eccentricity stellar orbits to be
\begin{equation}
\begin{split}
\frac{a_{\mathrm{critical}}}{a_{\mathrm{stars}}}=(1.60\pm0.04)+(4.12\pm0.09)\left(\frac{M_2}{M_1}\right) \\
+(-5.09\pm0.11)\left(\frac{M_2}{M_1}\right)^2
\end{split}
\end{equation}
when $M_{\mathrm{2}}$ / $M_{\mathrm{1}}$ is no greater than 0.5. For higher mass ratios we adopt the value of $a_{\mathrm{critical}}$=2.37 from \cite{stable89}.
\par No planet within the region of stability reaches an $\eta$ value much greater than 0.01 (1\% variations), which would imply that their atmospheres are similar to that of a single-star planet, with only slight changes in temperature for small periods of time. We do not expect that these systems will exhibit circulation differences, and therefore make the overarching conclusion that the atmospheres of circumbinary planets exhibit no strong or noticeable effects due to their unique irradiation patterns.
\subsection{Introduction of Obliquity} 
Seasons are an additional way in which we could produce variations over the single-star case. When orbiting two stars, the relative strength of a given season depends on which star is closest to the planet for the longest amount of time during that season. Therefore, we could expect hot summers and cold summers in a system with a low mass ratio. This is something not seen on a planet with only one star, so we investigate if the addition of a seasonal pattern (implemented by adding an axial tilt) can serve to excite large differences in a circumbinary planet's atmosphere. 
\par We introduce seasons in the typical way \citep{ClimateBook} where the daily averaged stellar irradiation at a given latitude varies with time as
\begin{subequations}
\begin{equation}
\bar{S}(\phi,t)=\frac{F}{\pi}[H(t)\sin\varphi\sin\delta(t)+\sin H(t)\cos\varphi\cos\delta(t)]
\end{equation}
\begin{equation}
\sin\delta(t)=\sin\psi\sin\lambda(t)
\end{equation}
\begin{equation}
\cos H(t)=-\tan\varphi\tan\delta(t),
\end{equation}
\end{subequations}
with $F$ the calculated flux, based on the location of the stars relative to the planet (see Section \ref{SandA_EBM}); $H$ is the hour angle at sunset; $\varphi$ is the latitude; and $\delta$ is the declination of the substellar point. Because there are two substellar points, one for each star, $\delta$ is further defined as the point directly in between the two substellar points - though the separation of the stars has been found to be negligible regardless. Further, $\psi$ gives the obliquity (tilt of the orbital axis); and $\lambda$ represents the angle of the planet's orbit relative to some standard `zero' point.
\par
\begin{figure*}
\epsscale{1.2}
\plotone{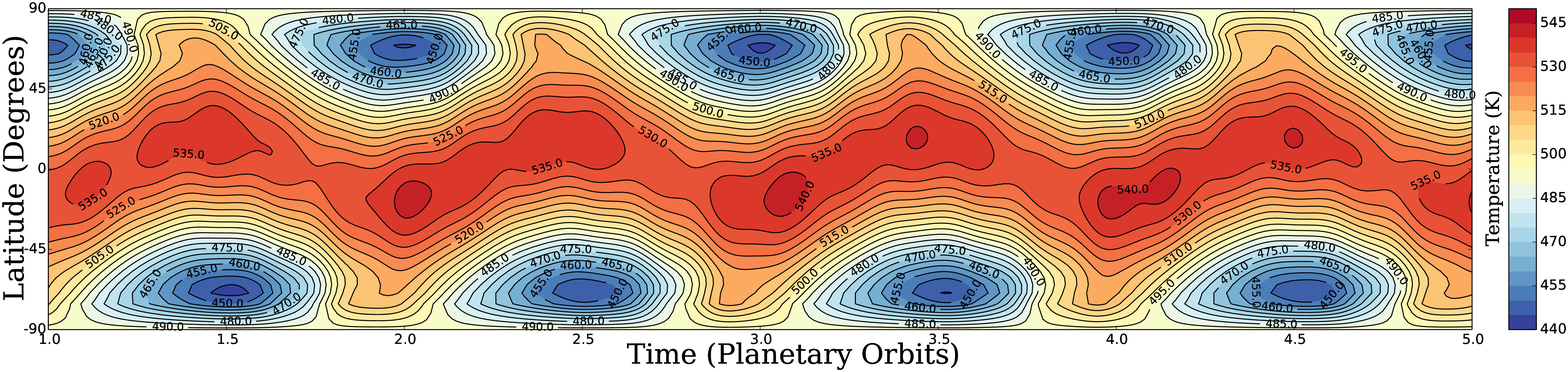}
\caption{Temperature map for Kepler 47b with a 30$^{\degree}$ obliquity. The typical seasonal pattern is present, with warm summers in the northern hemisphere while there is a cold winter in the southern hemisphere at the same time and vice-versa. While each summer/winter pattern is slightly different than the previous, this is not a strong effect. We find no additional variation over the single-star case as compared to the zero obliquity models.}
\label{seasonstemp}
\end{figure*}

 As before, we start by looking at the planet Kepler 47b. Using an obliquity of 30$^{\degree}$, and running this through the EBM, we begin to see the seasonal variation we expect with hotter summers in one hemisphere while there is a colder winter in the other hemisphere. Figure \ref{seasonstemp} shows the results for this case. We see no strong difference in the general seasonal pattern as compared to an equivalent single-star case with the same axial tilt.
\par Setting up the hypothetical models as before, with the same grid of stellar and planet orbital properties, we now vary the planetary obliquity between 0$^{\degree}$ (no axial tilt, same as before) and 90$^{\degree}$ (planet rotating completely on its side). In Figure \ref{Grid} we show results for obliquities of 15$^{\degree}$, 45$^{\degree}$, and 75$^{\degree}$ with the zero obliquity case subtracted off in order to focus only on the additional temperature variations caused by the seasons.
Compared to the zero obliquity case shown in the top left of Figure \ref{Grid}, we see the most additional variation for planets close in to any type of binary system (left region of subplots in Figure \ref{grid}). However, these planets are on unstable orbits (compare to the red line of stability in the 0$^{\degree}$ obliquity case), and are therefore unimportant to this discussion. The next region with significant additional differences are those planets far away from a equal mass binary system (top left of plots in Figure \ref{grid}). In this region, we expect planets to exhibit temperature patterns similar to their equivalent single-star case since they are far enough away from their hosts that the motions of the stars become less important as compared to their axial tilt. As with the zero obliquity case in Section \ref{grid}, we conclude that the atmospheres of circumbinary planets with any axial tilt are no different that their equivalent single-star case. Seasons are no more extreme for circumbinary planets than single-star planets.
\section{Conclusions}
\label{4_Conclusions}
\par We conclude that the temperature structures and wind patterns within the atmosphere of a circumbinary planet are negligibly different from its equivalent single-star case. This has important implications for future work, both with considerations of habitability and, more importantly to this work, continuing modeling effort of these planets.
\subsection{Modeling of Circumbinary Planets} 
\par We find a a maximum deviation in temperature of approximately 1\% for circumbinary planets on stable orbits. Based on our further modeling of Kepler 47b, a planet which experiences deviations of 0.2\%, we do not expect a deviation of 1\% to be large enough to lead to differences in circulation patterns. Therefore we conclude that the atmospheres of circumbinary planets, in all reasonably stable orbital configurations, are negligibly different from their equivalent single-star cases. Going forward, it is therefore reasonable to model circumbinary planets as their equivalent single-star case for studies of atmospheric circulation.
\subsection{Habitability of Circumbinary Planets} 
\par Although the ten known circumbinary planets are all of sufficient size to be considered gaseous, we can extend our results to make certain assumptions about the potential for habitability of moons or terrestrial planets within the habitable zone of binary star systems. We conclude that circumbinary planets should not be discounted as potential hosts for life, as their atmospheres should be of a similar nature to the single-star planets we know so well, though much more work can be done in order to study any possible circulation differences of circumbinary terrestrial planet atmospheres specifically. 
\subsection{Observables} 
\label{obs}
\par With our growing ability to observe planetary atmospheres, it becomes interesting to consider if we could measure the temperature variations of a circumbinary planet's atmosphere. During a secondary eclipse, we are able to determine the planet's thermal emission - emission which originates from the depth in the atmosphere which we have studied in this work. Ideally, by observing several different secondary eclipses and noting the positions of the stars relative to the planet at these points, we can predict the expected temperature differences between the two events and test the predictions based on the observations.
\par In reality, we do not expect these temperature variations to be large. For Kepler 47b, the maximum temperature variations between different points in orbit are of order a few Kelvin as predicted by the GCM. This is quite far out of the reach of current ground based efforts, which have errors of 100s of Kelvin \citep{Ground_1,Ground_2}. Recent best efforts to measure thermal emission from an exoplanet using Spitzer phase curves have achieved errors of only 20-60K \citep{Knutson09, Knutson12, Maxted13,Spitzer_Wasp14,Zellem14}, but this would still not be small enough to measure the extremely small temperature differences of a circumbinary planet with any certainty. While JWST will be even more powerful than Spitzer for atmospheric characterization \citep{JWST_Conf}, it is unlikely it will have the precision necessary (only a few Kelvin) in order to definitively say whether the expected small amplitude temperature variations are present in the atmospheres of circumbinary planets.

\bibliographystyle{apj}
\bibliography{apj-jour,REFS}

\end{document}